\documentclass{iopart}

\usepackage{graphicx}
\usepackage{dcolumn}
\usepackage{bm}
\usepackage[applemac]{inputenc} 
\usepackage{hyperref} 

\makeatletter
  \def\text#1{%
    \relax
    \ifmmode
      \mathchoice
        {\hbox{{\everymath{\displaystyle     }#1}}}%
        {\hbox{{\everymath{\textstyle        }#1}}}%
        {\hbox{{\everymath{\scriptstyle      }\let\f@size\sf@size\selectfont#1}}}%
        {\hbox{{\everymath{\scriptscriptstyle}\let\f@size\ssf@size\selectfont#1}}}%
      \glb@settings
    \else
      \mbox{#1}%
    \fi
  }
\makeatother

\graphicspath{{Figures/}} 

\begin{document}

\newcommand{\bsig}{{\bm \sigma}}
\newcommand{\btau}{{\bm \tau}}
\newcommand{\bmu}{{\bm \mu}}
\newcommand{\bS}{{\bm S}}
\newcommand{\bQ}{{\bf Q}}
\newcommand{\bk}{{\bf k}}
\newcommand{\bq}{{\bf q}}
\newcommand{\bM}{{\bf M}}
\newcommand{\bR}{{\bf R}}
\newcommand{\bt}{{\bf t}}

\newcommand{\bars}{\bar{s}}

\newcommand{\la}{\langle}
\newcommand{\ra}{\rangle}
\newcommand{\da}{\dagger}
\newcommand{\up}{\uparrow}
\newcommand{\down}{\downarrow}

\newcommand{\lc}{\lowercase}

\newcommand{\no}{\nonumber}
\newcommand{\be}{\begin{equation}} 
\newcommand{\ee}{\end{equation}}
\newcommand{\bea}{\begin{eqnarray}} 
\newcommand{\eea}{\end{eqnarray}}

\newcommand{\Q}{($0,0,\frac{2\pi}{c}$)}
\newcommand{\mul}{$\la\mu_{\text{loc}}^2\ra$}
\newcommand{\hch}{\hat{\chi}}

\def\bra#1{{\langle #1 \vert}}
\def\ket#1{{\vert #1 \rangle}}
\def\x#1{{\sigma_{#1}^{x}}}
\def\z#1{{\sigma_{#1}^{z}}}
\def\y#1{{\sigma_{#1}^{y}}}


\title{Moment screening in the correlated Kondo lattice model}

\author{M Siahatgar$^{1}$, B Schmidt$^{1}$, G Zwicknagl$^{2}$ and P Thalmeier$^{1}$}
\address{$^{1}$Max Planck Institute for Chemical Physics of Solids, 01187 Dresden, Germany}
\address{$^{2}$Technical University Braunschweig, 0531 Braunschweig, Germany}
\ead{bs@cpfs.mpg.de}
\begin{abstract}
The magnetic correlations, local moments and the susceptibility in the
correlated 2D Kondo lattice model at half filling are investigated.
We calculate their systematic dependence on the control parameters
$J_K/t$ and $U/t$.  An unbiased and reliable exact diagonalization
(ED) approach for ground state properties as well as the finite
temperature Lanczos method (FTLM) for specific heat and the uniform
susceptibility are employed for small tiles on the square lattice.
They lead to two major results: Firstly we show that the screened
local moment exhibits non-monotonic behavior as a function of $U$ for
weak Kondo coupling $J_K$.  Secondly the temperature dependence of the
susceptibility obtained from FTLM allows to extract the dependence of
the characteristic Kondo temperature scale $T^*$ on the correlation
strength U. A monotonic increase of $T^*$ for small U is found
resolving the ambiguity from earlier investigations.  In the large U
limit the model is equivalent to the 2D Kondo necklace model with two
types of localized spins.  In this limit the numerical results can be
compared to those of the analytical bond operator method in mean field
treatment and excellent agreement for the total paramagnetic moment is
found, supporting the reliability of both methods.
\end{abstract}


\section{Introduction}
\label{sec:introduction}

The understanding of strong correlations in mixed valent and heavy 
f-electron compounds is mostly based on two generic models described by the Anderson lattice Hamiltonian or, in the special case of almost integer valence, by the Kondo lattice Hamiltonian~\cite{hewson:93,newns:87}.
The latter model which will be the subject of this work represents an 
extreme limit where correlations in the f-orbitals are taken as infinitely large due to the Coulomb integral $U_f\gg t$ with  t denoting the hopping energy of conduction (c) electrons. On the other hand the Coulomb interaction $U$ of conduction electrons (and also the one between c- and f-electrons) is completely neglected in this model. The $U=0$ Kondo and Anderson models have the advantage of a simple and meaningful mean field solution~\cite{newns:87,bickers:87} with the constraint of only singly occupied f orbitals implemented by an auxiliary boson in mean field approximation. In the lattice this leads to hybridized quasiparticle bands with an exponentially reduced hybridization gap. Close to the half filled case with one c- and f- electron per site they have an effective mass $m^*\gg m_b$ much larger than the conduction electron band mass $m_b$~\cite{hewson:93}. The mean field Kondo lattice model may be merged with local density band structure calculations in the renormalized band theory~\cite{zwicknagl:92} leading to a powerful method to calculate realistic Fermi surfaces for Ce and Yb heavy fermion compounds, including crystalline electric field effects.

It has been known that a strong Coulomb repulsion among the conduction electrons significantly alters the electronic properties of a metal. First and foremost, the kinetic energy as measured by the (effective) band width will be reduced and eventually vanish at a metal-to- insulator transition. Second, the trend towards localization is accompanied by the appearance of magnetic correlations.

The central focus of the present paper is the influence of the above-mentioned correlation effects on the screening of local moments. Of particular interest is the question how the Kondo energy scale is affected by conduction electron correlations. We adopt a two-dimensional Hubbard model at half-filling for the conduction electrons where correlation effects have been found to affect strongly the electronic properties. As the interacting conduction electron Hamiltonian, i. e., the two-dimensional Hubbard model cannot be solved exactly for an infinitely extended system we extract the relevant information from suitably chosen finite clusters. This approach seems justified considering the local character of the quantities to be investigated. The relevance of the cluster approach is assessed by comparing its results to the predictions of a constrained mean-field theory for an infinite system in the limit of extremely strong Coulomb repulsion.

Attempts to include the effect of correlations between conduction electrons which may become important when the latter originate from d-orbitals have sofar been mostly limited to the Kondo impurity models using various analytical techniques like perturbation theory~\cite{khaliullin:95,neef:03} and 1/$N_f$ expansion for small U, a Schrieffer-Wolff approach~\cite{schork:94} and RVB ansatz~\cite{khaliullin:95} for large U, a scaling approach~\cite{li:95}  and also numerical techniques like NRG~\cite{takayama:98}. In the impurity problem it was concluded that the Kondo energy scale $T^*$ increases with U~\cite{khaliullin:95,neef:03}.

Concerning concentrated systems with magnetic moments at every lattice site the situation is rather controversial. In the case of non- interacting conduction electrons (U=0) the electronic properties will be determined by the competition between the energy gain due to (local) Kondo singlet formation and to developing (long-range) magnetic correlations. It is to be expected that the subtle balance between these two tendencies will be affected by conduction electron repulsion.

The  correlated Anderson lattice problem was treated with a Gutzwiller variational method~\cite{itai:96} and the 1D  correlated Kondo lattice with DMRG approach~\cite{shibata:96}. It was further investigated with QMC simulations and a fermionic bond operator method~\cite{feldbacher:02} used before in the uncorrelated case~\cite{jurecka:01}.
The strongly correlated ($U\gg t$) 'Kondo necklace' limit of the Kondo lattice was treated using exact diagonalization~\cite{zerec:06a}.
It was suggested~\cite{itai:96} that the Kondo scale actually decreases for increasing U, in contrast to the impurity results.
The competition of singlet formation and induced inter-site RKKY coupling and its signature in thermodynamics was studied in Ref.~\cite{zerec:06b} using ED and FTLM methods for finite clusters of the 2D uncorrelated Kondo lattice and in Ref.~\cite{igarashi:95} for 1D chains in the strongly correlated limit.

The present work which uses the unbiased numerical ED and FTLM methods as well as the analytical bond operator technique has two clear objectives:  i) Firstly  to investigate the characteristic T=0 on-site and inter-site correlations and local  paramagnetic moment \mul~  systematically in the whole ($J_K,U$) parameter range of the correlated Kondo lattice model, where $J_K$ is the local antiferromagnetic exchange coupling. In particular the total local paramagnetic moment  is an excellent measure for the formation or breakup of the Kondo singlet state as function of control parameters~\cite{zerec:06b}. We will show that it exhibits an unexpected non-monotonic behaviour that has not been reported before.
ii) Secondly we calculate the finite temperature behaviour of susceptibility and specific heat with FTLM  on a small tile of the square lattice. From the maximum position we may extract the characteristic 'Kondo' temperature scale $T^*$ and in particular we investigate its systematic dependence on the correlation strength U. We will resolve the ambiguity mentioned before and show that  $T^*$ increases monotonically with U. The ED and FTLM calculations will be performed on small Kondo clusters with open boundary conditions  which have a more realistic one-particle density of states than the one with periodic boundary conditions~\cite{zerec:06b}.

We stress from the outset that in the context of the FTLM calculations on a finite cluster the meaning of $T^*$ is that of an average energy of low energy singlet-triplet excitations obtained from the maximum in the temperature dependent susceptibility and specific heat. This is indeed the way how an estimate of $T^*$ in a real concentrated Kondo compound may be  obtained. It is not the same as the genuine single ion  Kondo temperature $T_K$ in the continuum limit which is exponentially small compared to the hopping energy. This difference between lattice and single ion Kondo scales appears already within approximate analytical theories~\cite{newns:87}. 

Nevertheless a comparison of ED numerical results for finite clusters with analytical results obtained by the bond operator method for the extended lattice and in the large U limit is useful and legitimate. Since the former are exact it  allows one to judge how far the large U model approximation is justified and give an estimate of the reliability and accuracy of the bond operator approximation.
This theory has previously been used for quantum critical properties of the localized model in the large U limit~\cite{zhang:00,langari:06,thalmeier:07}. A generalized fermionic version has also been applied to the finite U case~\cite{jurecka:01}. In this comparison we focus on the paramagnetic local moment size and on-site correlations as function of U and on the finite temperature susceptibility.

We note that  the investigation of finite Kondo atom clusters on nonmagnetic surfaces is also of potential experimental interest.
Sofar single  magnetic Co adatoms embedded in metallic Cu$_n$ clusters~\cite{neel:08} and two site Co-Co Kondo clusters
connected by Cu$_n$ chains~\cite{neel:11} have been investigated. However in these cases of a metallic substrate
it is the simple conduction electron density of the substrate surface that plays the dominant role for the Kondo properties.
The present investigation would be relevant for correlated Kondo clusters weakly adsorbed on an inert insulating substrate without chemical bonding of substrate and Kondo atoms.  This has, to our knowledge, not been experimentally realized until now.
Nevertheless it is clear that the investigation of finite size Kondo clusters is of great interest in itself and not only in view of the 
bulk Kondo lattice materials.

%
\begin{figure}
\centering
\includegraphics[width=75mm]{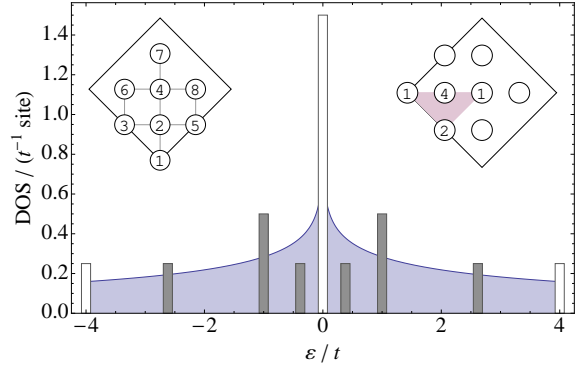}
\caption{Single particle DOS for the 2D n.n. tight binding model in the continuum limit (full line) with logarithmic singularity at
Fermi level for half filling ($\epsilon =0$). Number of states for eight site cluster is shown for OBC (full bars) and 
PBC (open bars). Left inset: Eight site cluster: Right inset : \bk - points for PBC.}
\label{fig:Fig1}
\end{figure}
%

\section{Model definition and single particle spectrum}
\label{sec:model}
 
The correlated Kondo lattice model ,sometimes called Kondo-Hubbard
(KLU) model  contains three terms.  As in the non-interacting model there is
the hopping term for conduction electrons $c_{i\sigma}$ ($\sim t$) and
the Kondo exchange term ($\sim J_K$) to localized spins $\bS_i$ at
every site.  Their localization may be considered as result of their
large repulsion $U_f$.  As mentioned before the Coulomb repulsion $U$
for conduction electrons is usually neglected in the lattice model,
although well studied for the Kondo impurity case.  It is, however,
very interesting to include its effect, both because it is physically
present, in particular in intermetallic 3d-4f compounds and also
because it allows to study theoretically the continuous crossover from
the metallic Kondo screening case to the insulating case of
interacting local spin dimers.  The model is given by
\be
H_{\text{KLU}}=-t\sum_{i,j;\sigma}c^\dagger_{i\sigma}c_{j\sigma}
+U\sum_i n_{i\uparrow}n_{i\downarrow}
+J_{\text K}\sum_i\btau_i\cdot\bS_i
\label{eq:KLU}
\ee
where $n_{i\sigma}=c^\dagger_{i\sigma}c_{i\sigma}$.  In 2D the first.  n.n.
hopping term leads to the single particle spectrum given by the
dispersion $\epsilon_\bk=-t(\cos k_x +\cos k_y)$.  In the continuum
limit ($N\rightarrow\infty$) it leads to a well known density of
states (DOS) function shown in Fig.~\ref{fig:Fig1} by the full line
which has a logarithmic singularity a $\epsilon =0$ corresponding to
half filling.  As was done in Ref.~\cite{zerec:06b} for the
uncorrelated KL model ($U=0$) we will use the exact diagonalisation
with Lanczos method for finite square lattice tiles.  For the
systematic classification of finite size tiles on the square lattice
we refer to~\cite{schmidt:11}.  Due to the large number of eight
states per site $\{S_z=\pm\frac{1}{2}\} \otimes\{ n=0; n=1
(\tau_z=\pm\frac{1}{2}); n=2\}$ in the KLU model only tiles with eight
sites can be used (see inset of Fig.~\ref{fig:Fig1}).  This precludes
also finite size scaling such as may be performed in the 1D case~\cite{tsunetsugu:92,shibata:96}.
Furthermore as in Ref.~\cite{zerec:06b} we restrict to the half filled case
with the number of conduction electrons given by $n_c/N=1$ (N=8). Due to  small cluster size (N)
$n_c$ can have only discrete values and therefore the next possible value $n_c/N= 3/4$ 
in the singlet spin sector is already far from the interesting half filled case.

As discussed before~\cite{zerec:06b} one may choose both open and
periodic boundary conditions for the tiles (OBC and PBC respectively).
Their single particle spectrum differs greatly.  While for PBC most of
the states are put directly at the Fermi energy, giving little
resemblance to the continuum DOS the spectrum is more realistic for
OBC with more even distribution of energies (Fig.~\ref{fig:Fig1}).  It
was shown in the case $U=0$~\cite{zerec:06b} that this leads to a
qualitatively different moment screening or singlet formation for
small $J_K$.  In this work we use exclusively OBC.  Naturally, since there is a
finite gap between occupied and unoccupied conduction electron states
for half filling (Fig.~\ref{fig:Fig1}), the true continuum behavior of
the infinite Kondo lattice cannot be literally described, the numerical ED results
presented in the following rather describe finite 'Kondo molecules'.
As mentioned in the introduction they may in fact be artificially
generated by adsorption of 3d clusters on nonmetallic surfaces.  On
the other hand we will focus mostly on the local moment and on-site or
n.n. correlations.  At least in the strong correlation limit ($U/t\gg
1$) it turns out that the local quantities obtained numerically for
the eight-site cluster agree rather well with the infinite lattice
results (Sec.~\ref{sec:bondop}).

%
\begin{figure}
\centering
\includegraphics[width=85mm]{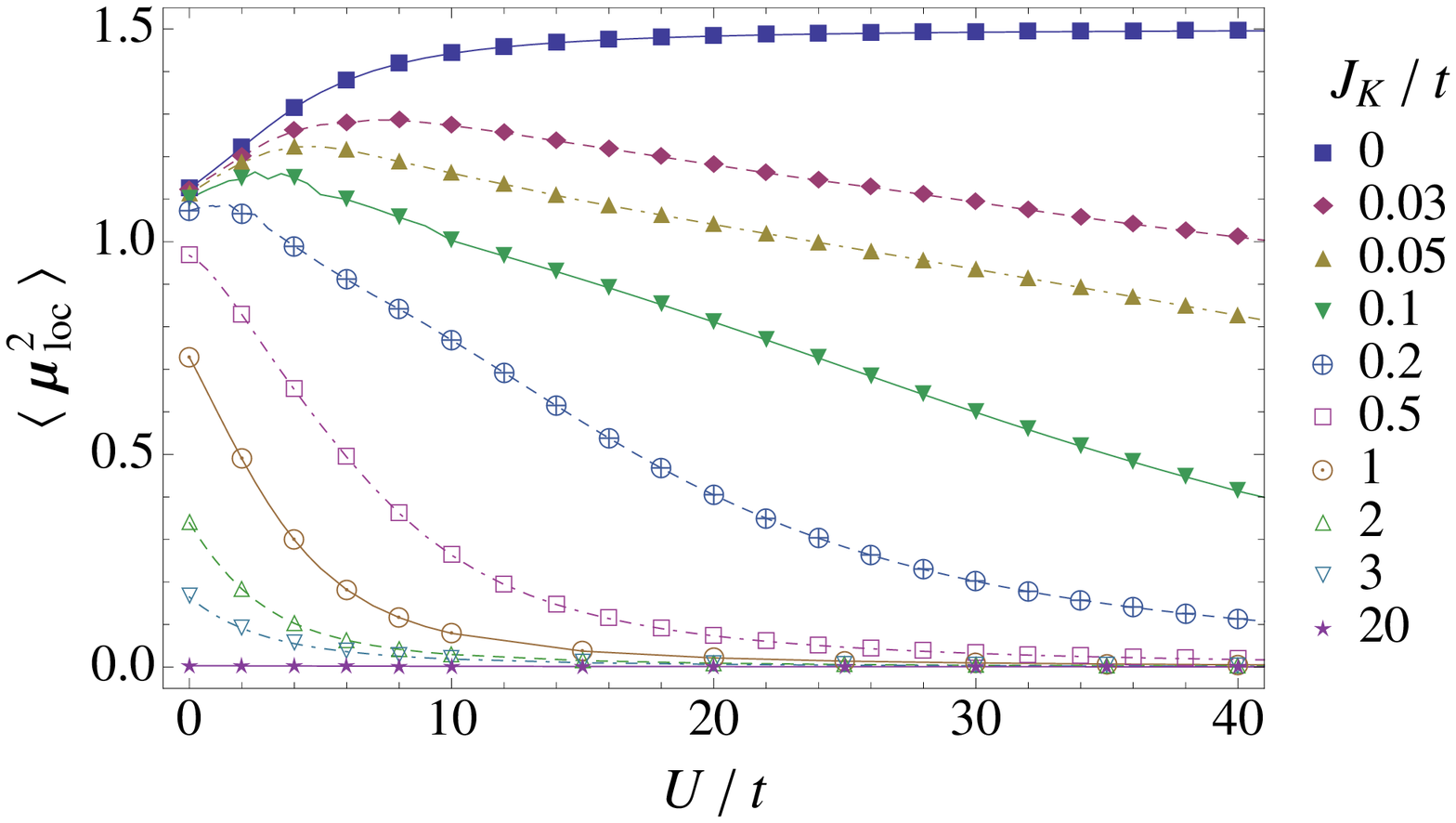}
\includegraphics[width=85mm]{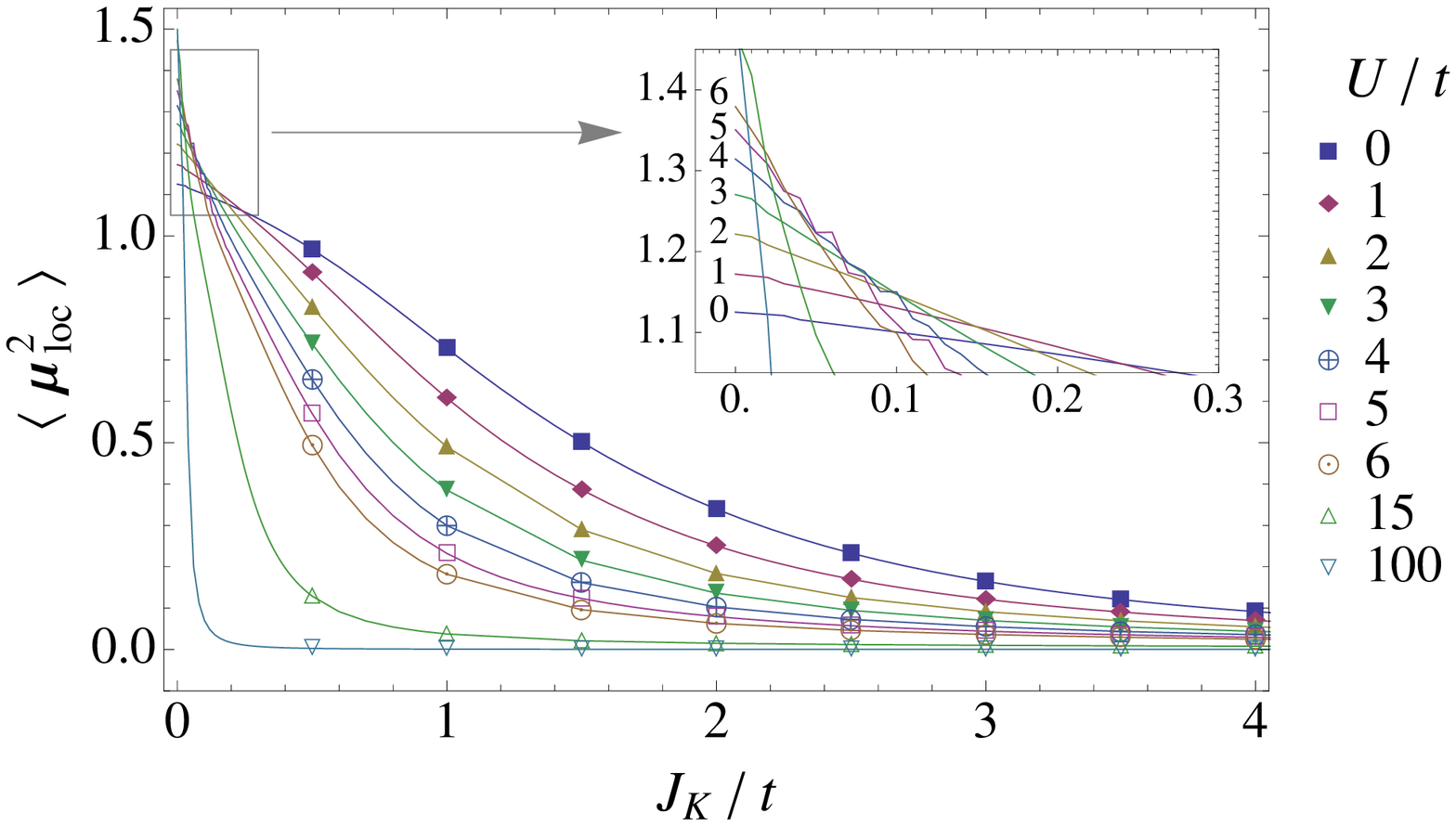}
\caption{(a) Effective paramagnetic total moment squared as function
of U/t for various $J_K/t$.  A flat maximum of the moment appears for
weak Kondo coupling.  (b) Corresponding varation of
$\langle\mu_{\text{loc}}^2\rangle$ as function of $J_K/t$ for
different $U$.}
\label{fig:Fig2}
\end{figure}
%

\section{Numerical determination of the local moment and thermodynamic properties}
\label{sec:ED}

To determine the ground state and thermodynamic properties of our
system, we diagonalize the Hamiltonian iteratively on a small tile,
applying the Lanczos algorithm.  We also use the low-lying states and
their energies to construct thermodynamic expectation values, in
particular to calculate the partition function and its
derivatives~\cite{jaklic:00}.  We stress that this unbiased method is exact and
reliable for finite clusters and can therefore be used as a reference to compare with other numerical
or analytical methods.
We note that the large number
of (eight) states per site prevents the use of larger clusters and 
the application of a finite size scaling procedure such as
described in detail in Ref.~\cite{schmidt:11} for a
two-dimensional $S=1/2$ model with just two states per site.
Therefore using the Lanczos method for the present model
is limited to tiles with eight sites (a maximum of 10 sites seems
attainable).  Even then the largest subspace with total spin
$S^{\text{tot}}_z=0$ has dimension $739\,162$ ($38\,165\,260$ for
$N=10$ sites).  Nevertheless, as the comparison with analytical
results will show, the local moment and on-site correlations
for the small tile give a realistic approximation to the extended
Kondo lattice behavior.

The total paramagnetic local moment is given by
\begin{equation}
   \langle \mu_{\text{loc}}^{2}\rangle
     =
    \left\langle\bS_{i}^{2}\right\rangle+
    \left\langle\btau_{i}^{2}\right\rangle+
    2\left\langle\bS_{i}\cdot\btau_{i}\right\rangle.
    \label{eq:muloc}
\end{equation}
It contains the spin fluctuations of both the itinerant and localized
spins and their antiferromagnetic on-site correlations due to the Kondo
coupling.  This is a central quantity for investigating the influence
of correlations on the Kondo effect because the Coulomb repulsion $U$
tends to localize the $\btau$ spins and thus influences strongly their
singlet formation with $\bS$ spins.  For the uncorrelated case ($U=0$)
it was shown previously that the on-site Kondo coupling also induces
effective RKKY inter-site coupling between the localized spins.  This
mechanism is still present for non-zero $U$ but it has to compete with
the effective superexchange.  Therefore, in addition to the total
moment we will also investigate the on-site singlet correlation
$S_{\text{KS}}=\langle\btau_i\cdot\bS_i\rangle$ and nearest neighbor
$(i,j)$ correlation function $S_{ij}= \langle\bS_i\cdot\bS_j\rangle$.
Due to the smallness of the cluster it is only reasonable to calculate
it up to next nearest neighbors. The same would be true for the 
correlation function $\langle\bS_i\cdot\btau_j\rangle$ characterizing
the 'screening cloud' of a given localized spin $\bS_i$ which is not evaluated here
because its characteristic length scale is much larger than the cluster size.
In the case of a single Kondo impurity in the continuum limit it extends over a range of $\xi=v_F/T_K$ where
$v_F$ is the Fermi velocity and $T_K$ the single ion Kondo temperature. Note however 
that even in this simple case  the screening cloud and its length scale $\xi$ have not yet
been observed in reality~\cite{affleck:09}.

With the FTLM approach~\cite{jaklic:00} it is possible to
calculate the temperature dependence of various important
thermodynamic quantities like specific heat $C(T)=N_Ak_B\hat{C}(T)$ and uniform
susceptibility $\chi(T)=N_A\mu_0(g\mu_B)^2\hch(T)$ per mole as cumulants of simple operators,
e.\,g.,
\bea
    \label{eq:FTLM}
    \hch(T)&=&
    \frac{1}{N}
    \frac{1}{k_{\text B}T}
    \left[
    \left\langle
    \left(S_{z}^{\text{tot}}\right)^{2}
    \right\rangle_T
    -\left\langle S^{\text{tot}}_z\right\rangle_T^2
    \right],
    \no
    \\
    S_{z}^{\text{tot}}&=&
    \sum_{i=1}^{N}\left(S_{i}^z+\tau_{i}^z\right),
    \\
    \hat{C}(T)&=&
    \frac{1}{N}
    \frac{1}{\left(k_{\text B}T\right)^2}
    \left[
    \left\langle
    \left(H_{\text{KLU}}\right)^{2}
    \right\rangle_T
    -\left\langle H_{\text{KLU}}\right\rangle_T^2
    \right],
    \nonumber
\eea
where $N_{\text A}$ is the Avogadro number, $\mu_{0}$ is the magnetic
permeability, $\mu_{\text B}$ and $k_{\text B}$ are the Bohr magneton
and the Boltzmann constant, $g$ is the gyromagnetic ratio, and $N$ is
the size of our tiles (or number of sites).  For simplicity, we assume
the same gyromagnetic ratio for both the $\bm\tau$ and the $\bm S$
spins.  The thermal expectation values are to be calculated with the
statistical operator $\rho(T)=\exp(-\beta H_{\text{KLU}})$, $\beta
=1/\left(k_{\text B}T\right)$.  In this work we restrict to zero
magnetic field, and as we are working with a finite system size $N$,
we can safely set $\left\langle S^{\text{tot}}_z\right\rangle_{T}=0$
in the first equation.  Numerically, for each set of Hamiltonian
parameters $\{t,U,J_{\text K}\}$, the calculation of thermal averages
involves two averaging procedures: Firstly, a random starting
wave function is chosen, and the Lanczos algorithm yields up to ${\cal
O}(100)$ extremal eigenvectors and eigenvalues.  Secondly between
$100$ and $500$ of these Lanczos runs with different starting
wavefunctions are performed.  All eigenvalues and eigenfunctions
obtained in this way are subsequently used to calculate the traces
over the statistical operator as indicated in Eqs.~(\ref{eq:FTLM}),
leading to the desired thermal expectation values.  For details we
refer to Ref.~\cite{jaklic:00}.

%
\begin{figure}
\centering
\includegraphics[width=85mm]{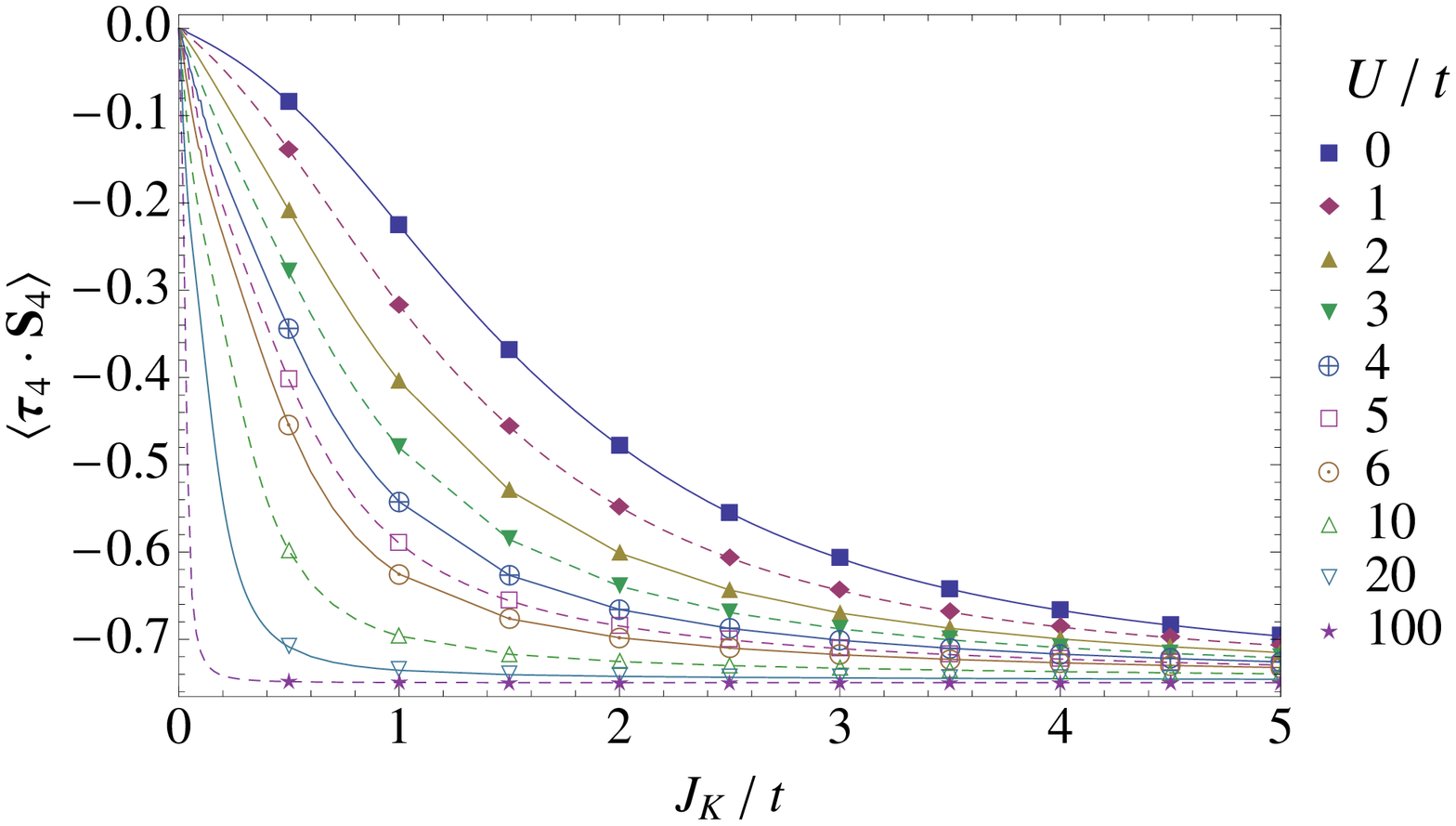}
\includegraphics[width=85mm]{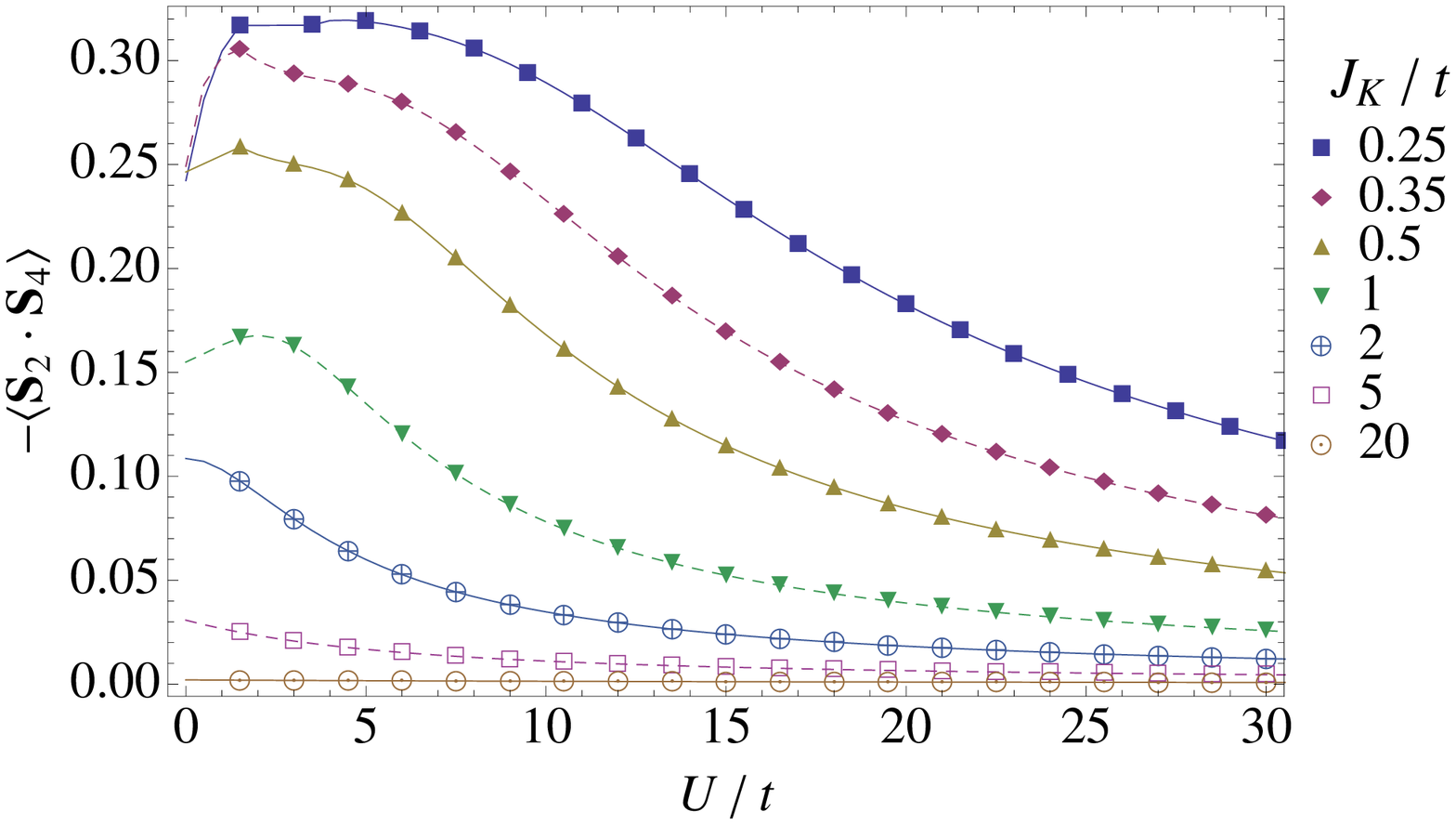}
\caption{(a) On-site Kondo-singlet formation between localized $\bS$ and itinerant $\btau$ spins as function of Kondo coupling 
for various strengths of Coulomb repulison. (b) Nearest neighbor AF correlations between localized spins as function of U/t for various
Kondo coupling strengths. For U=0 the n.n. coupling is of induced RKKY type while for larger U it is dominated by the superexchange term.}
\label{fig:Fig3}
\end{figure}
%

Although we do not consider the magnetization $M$ (per site) of the
correlated Kondo model explicitly, we may get some information about
finite-field properties by calculating the third-order susceptibility
defined through the expansion 
\bea
M=\chi\cdot B +\frac{1}{3!}\chi^{(3)}\cdot
B^3+\dots
\label{eq:magnet}
\eea
for an applied magnetic field $B$.  Here 
$\chi^{(3)}(T)=(N_A\mu_0)(g\mu_B)^4\hat{\chi}^{(3)}(T)$
is given by a higher order
cumulant according to
\bea
    \hat{\chi}^{(3)}(T)=
    \frac{1}{N}
    \frac{1}{\left(k_{\text B}T\right)^3}
    \left[
    \left\langle
    \left(S_{z}^{\text{tot}}\right)^{4}
    \right\rangle_T
    -3\left\langle
    \left(S^{\text{tot}}_z\right)^2
    \right\rangle_T^2\
    \right]
    \label{eq:thirdsus}
\eea
This quantity is a measure of the nonlinearity of magnetization at low
field strength $B$. It has been discussed previously for a localized
spin model~\cite{schmidt:05} and plays a significant role in the
discussion of some heavy fermion compounds~\cite{ramirez:92}.

\section{Discussion of numerical results}
\label{sec:discussion}

First we discuss results for the total local paramagnetic  moment presented in Fig.~\ref{fig:Fig2}(a). Two counteracting trends determine its size: On one hand the increase in U localizes the $\btau$ -spins and leads to an increase of $\langle\btau_i^2\rangle$ from $3/8$ for $U=0$ to $3/4$ for $U/t\gg 1$. On the other hand for finite $J_K$ the Kondo term establishes the AF on-site singlet correlation $\langle\btau_i\cdot\bS_i\rangle < 0$. Therefore in the case $J_K$=0 the moment increases monotonically from $9/8=1.125$  to $2\cdot(3/4)=1.5$ while for any finite $J_K$ in the large U limit when $\btau_i$ becomes localized the moment will decrease due to singlet formation. For moderate J$_K$ there is an initial increase of   $\la\mu_{\text{loc}}^{2}\ra$  with $U$ due to the first correlation effect and eventually a decrease due to the effect of $J_K$ when $\btau$ spins become localized at larger $U$. In between a maximum in   $\la\mu_{\text{loc}}^{2}\ra$ develops as function of $U$. For larger J$_K$ the singlet formation effect is so strong that it overwhelms the increase in  $\langle\btau_i^2\rangle$ and therefore no maximum appears beyond $J_K/t$=0.1. This effect is completely dominated by local correlations and should not be strongly influenced by the tile size.
The corresponding local moment dependence on $J_K$ for various U is shown in  Fig.~\ref{fig:Fig2}(b). For the uncorrelated $U=0$ case the singlet formation leads to a continuous decrease of $\la\mu_{\text{loc}}^{2}\ra$ with $J_K$. Increasing $U$ facilitates this formation due to the localization of $\btau$ spins and the decrease becomes progressively steeper as function of $J_K$. In the limit $U\rightarrow\infty$ an arbibtrary small $J_K$ will lead to the singlet ground state.

The singlet formation may also be monitored directly by the on-site correlation $\langle\btau_i\cdot\bS_i\rangle$ which is shown in  Fig.~\ref{fig:Fig3}(a). Starting from zero at $J_K=0$ it becomes increasingly antiferromagnetic for growing $J_K$ until the singlet value $-3/4$ is reached. Again the latter is approached more rapidly with increasing correlation strength $U$.
The nearest neighbor induced AF spin correlations are presented in Fig.~\ref{fig:Fig3}(b). For U=0 these correlations are of the induced RKKY type~\cite{zerec:06b}. They decrease with increasing $J_K$ because their evolution is impeded by the increasing singlet formation. For a fixed but small $J_K$  the antiferromagnetic inter-site correlation first increases for small U due to the reduction of single and double occupancies and for larger U it falls off again due to the reduction of the superexchange $4t^2/U$ with increasing U. For larger $J_K$ this effect is more pronounced as function of U.

A recurrent topic in previous analytical theories of the correlated Kondo impurity model is the U-dependence of the Kondo energy scale.
From a practical viewpoint it is frequently taken as the maximum position $T^*$ of magnetic specific heat or susceptibility which corresponds to an average singlet-triplet excitation energy.
In the present context of finite size tiles the $U$ or $J_K$ dependence of the characteristic  temperature scale $T^*$ can be conveniently obtained from the FTLM results according to Eq.~(\ref{eq:FTLM}) for both quantities. 
Although it is not identical to the single ion Kondo temperature $T_K$ of the impurity model one may expect a comparable qualitative dependence on  $U$ or $J_K$  may be comparable.

%
\begin{figure}
\includegraphics[width=75mm]{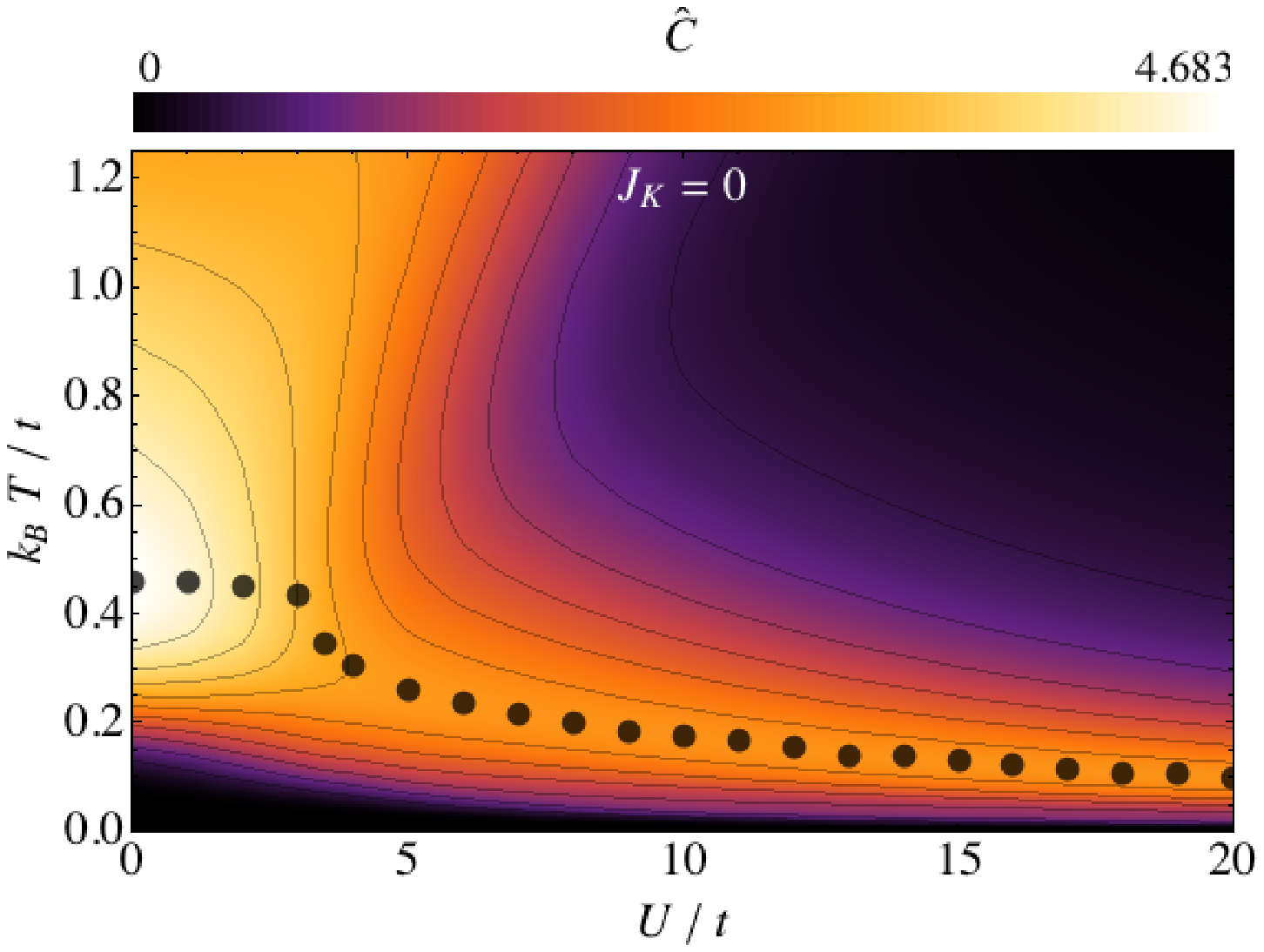}
\includegraphics[width=75mm]{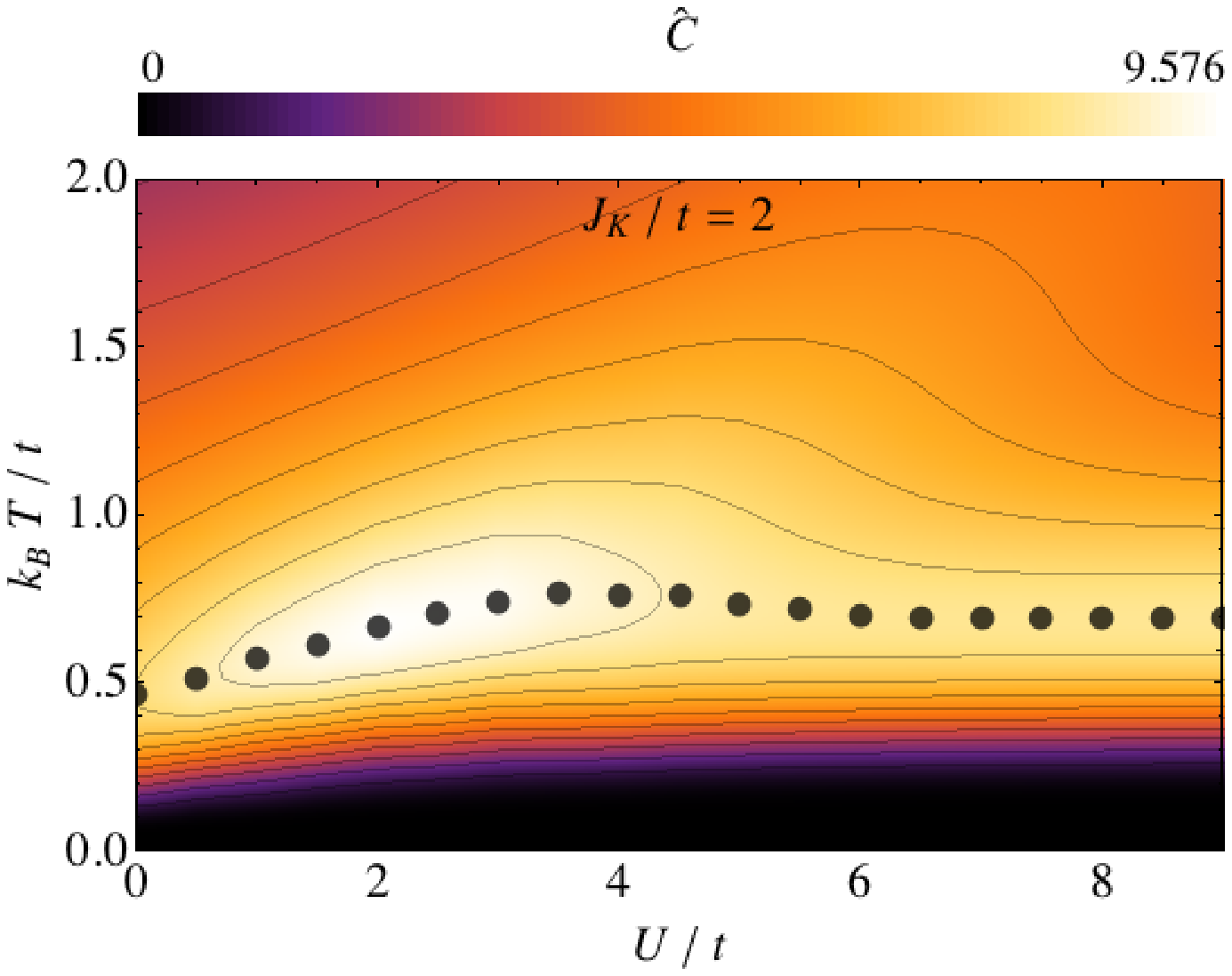}
\caption{Specific heat from FTLM as function of temperature and interaction strength U.
(a) for the eight site Hubbard tile ($J_K = 0$) the $C(T)=(N_Ak_B)\hat{C}(T)$ evolution as function of U shows the splitting
of charge and spin fluctuation peaks (upper and lower peaks respectively) for  $U>4t\approx W/2$ ( $W=8t $
is the total band width).
(b) for finite $J_K$ the U=0 peak  defines the Kondo temperature scale due to singlet formation. It increases with U until 
it reaches a plateau for $U>4t$. }
\label{fig:Fig4}
\end{figure}
%

The contour plot of $C_V(T)$ in the T-U plane is shown in Fig.~\ref{fig:Fig4}(a),(b) for $J_K/t =0,2$ respectively. The former corresponds to the case of the pure Hubbard model. It illustrates that the high temperature peak at $U=0$ due to uncorrelated charge fluctuations splits into a lower temperature spin fluctuation peak associated with $J=4t^2/U$ and a broad continuum at very high temperatures. The charge fluctuation peak at U=0 correspond to the excitation energy of $\approx 0.75t$ between the highest occupied and lowest unoccupied states around $\epsilon =0$ in Fig.~\ref{fig:Fig1}.
The behavior as function of $U$ may be viewed as a precursor of the thermodynamic Mott-Hubbard transition for a finite tile size. 
When $J_K$ is turned on the specific heat peak is dominated by the lowest singlet triplet excitation energies. It increases linearly for small $U$ (Fig.~\ref{fig:Fig4}(b)) and reaches a plateau around $U\approx 6t$ corresponding to the highest excitation energies in Fig.~\ref{fig:Fig1}.

\begin{figure}[Ht]
\includegraphics[width=75mm]{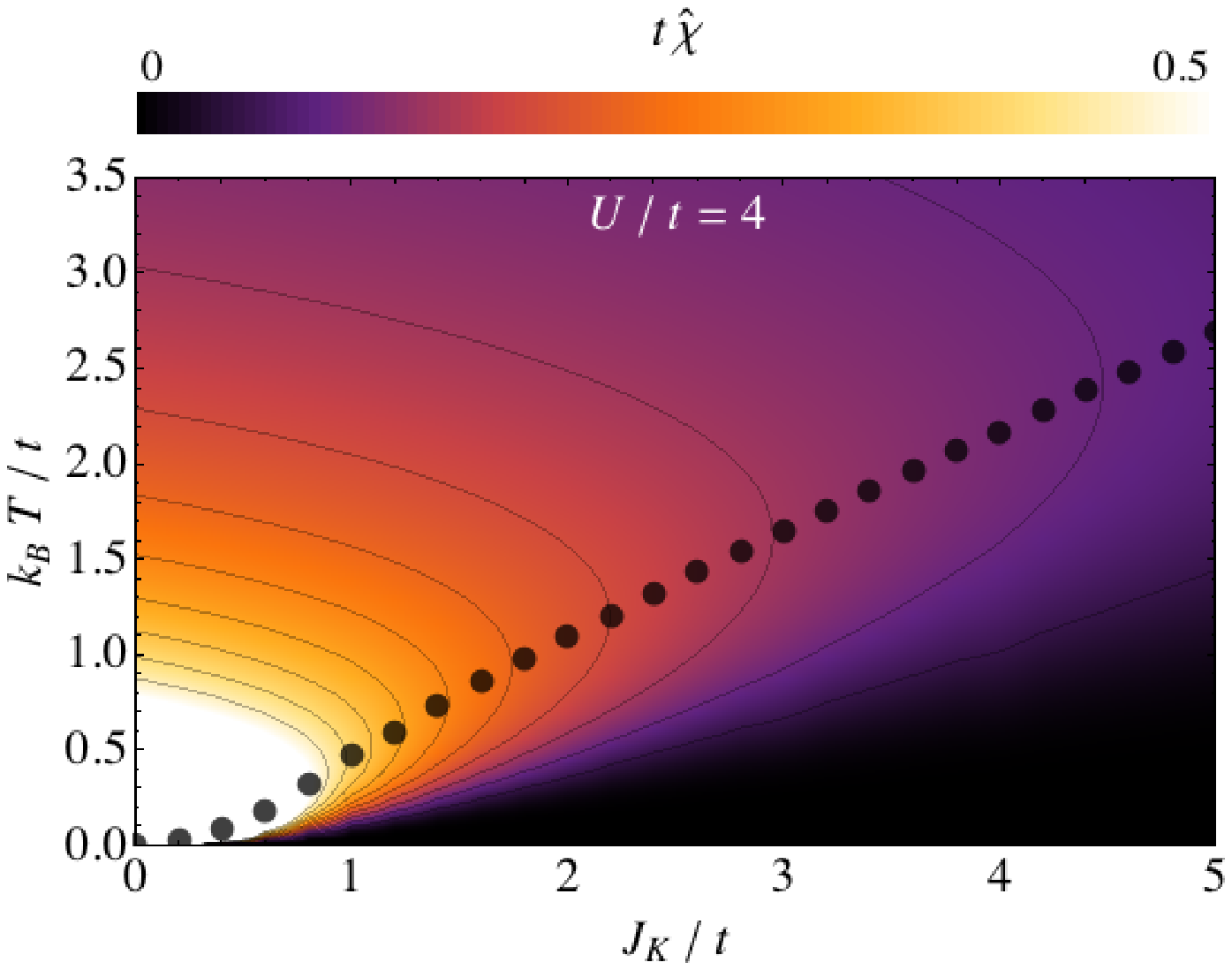}
\includegraphics[width=75mm]{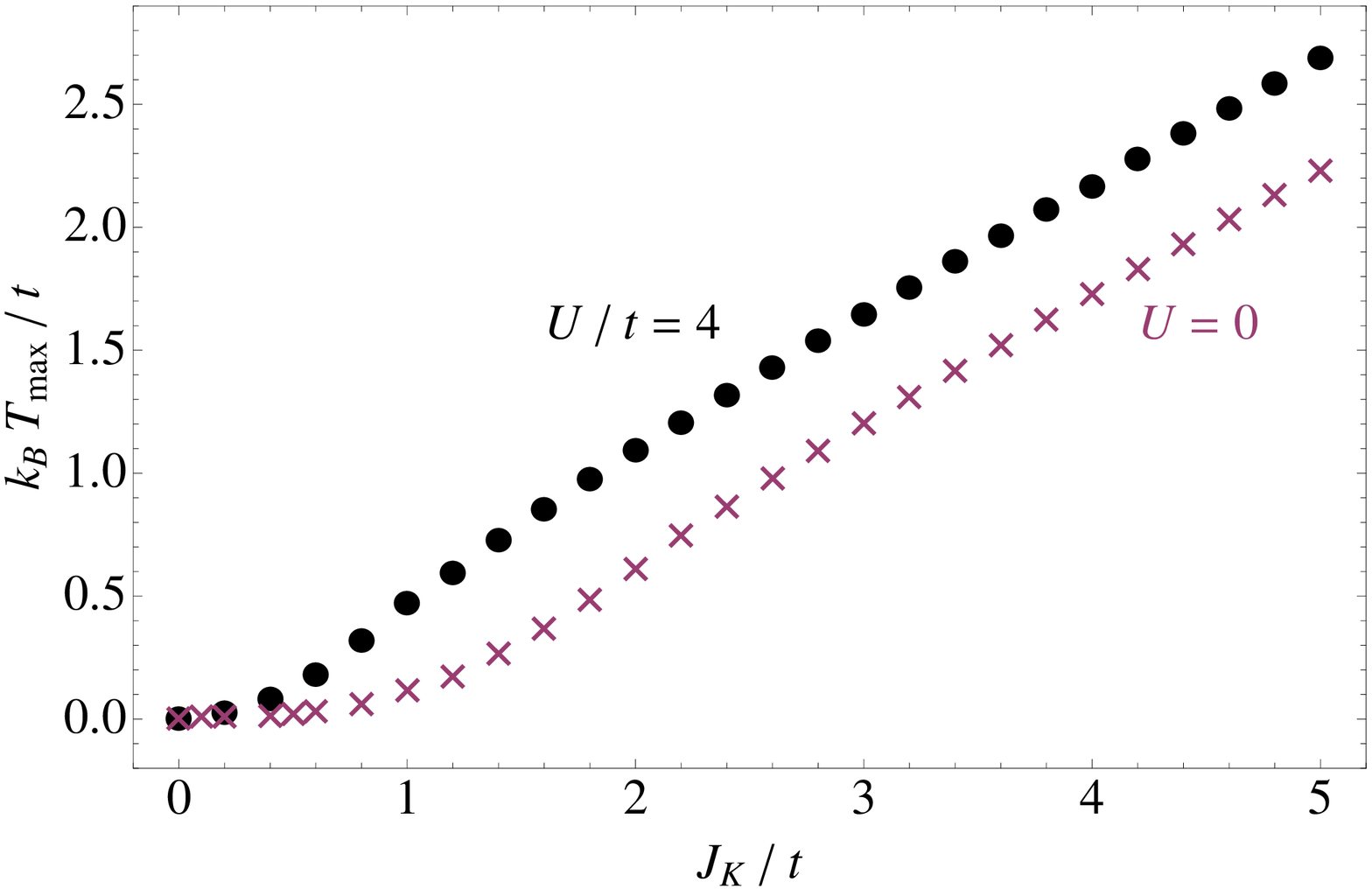}
\caption{Susceptibility from FTLM as function of temperature and Kondo coupling $J_K$.
(a) Contour plot of dimensionless $t\hch(T,J_K)$. The maximum (black dots) defines the Kondo temperature scale $T^*$  of singlet formation for $U/t=4$. It increases monotonically with $J_K$. 
(b) For comparison $T^*$ for the noninteracting $(U=0)$ case is shown by white crosses demonstrating the 
increase of $T^*$ with $U$.}
\label{fig:Fig5}
\end{figure}

To get rid of the influence of charge fluctuations we also calculated the spin susceptibility with FTLM. The results are presented in
 Figs.~\ref{fig:Fig5}(a),\ref{fig:Fig6} as function of $J_K,U$ respectively. In the former a contour plot shows the evolution of susceptibility maximum or Kondo temperature scale $T^*$ (black dots) as function of $J_K$ for constant $U/t=4$. For comparison the $T^*$ maxima for the uncorrelated $U=0$ Kondo lattice tile are also given in Fig.~\ref{fig:Fig5}(b) by the crosses. In both cases $T^*$ increases linearly with $J_K$ in the strong coupling limit $J_K/t\gg 1$. For small $J_K$ the values for $U=0$ are considerably below the results for finite U which indicates the increase of the Kondo temperature scale $T^*$ with U. This is also seen directly in  Fig.~\ref{fig:Fig6} where $T^*$ (black dots) can be seen to increase linearly with U for small U and reach a plateau for $U > 6t$ similar as for the specific heat peak before. In that figure we also included the peak position of the third order susceptibility  $\hat{\chi}^{(3)}(T)$ (white dots) which also increases with U. This quantity is experimentally accessible\cite{ramirez:92}. It peaks at a systematically lower temperature than the first order susceptibility. The reason is that it characterizes the strongest nonlinear increase of the local moment with field which happens precisely in the temperature region where the screened local moment and susceptibility at zero field drops to zero (see Fig.~\ref{fig:Fig6}). For comparison the size of the spin gap from analytical calculation in the large U limit (Eq.~(\ref{eq:spingap})) shown by white crosses is seen to be sandwiched between the above FTLM values for the first order (black dots) and third order (white dots) susceptibility peak positions.
 
%
%

\section{Bond operator treatment of the strongly correlated 'Kondo necklace' limit}
\label{sec:bondop}

A deeper insight into the phases and excitations of low dimensional quantum magnets requires the application of both numerical and analytical techniques. As demonstrated e.g. by the spin wave analysis of the frustrated 2D spin systems ( Ref.~\cite{schmidt:11} and references cited therein) analytical results for the extended system, even if approximate or only available in limiting cases , are very helpful to understand the systematics of numerical ED results for finite clusters. It is therefore perfectly legitimate to proceed in a similar way for Kondo lattice models. Because we focus on the paramagnetic phase we  will, however, use the bond operator approach as the appropriate analytical technique for comparison in the large U limit.

In the following we will derive such analytical results in the limit of large conduction electron correlations ($U\gg 2zt$, z=4 is the coordination number) at half filling ($n_c=1$). In this case the  charge fluctuations may be eliminated from the hopping and Hubbard terms leading to a pure exchange term of the now also localized conduction electron spins $\btau_i$. Thus the appropriate large U limit of the model in Eq.~(\ref{eq:KLU}) is a localized 2D spin Hamiltonian~\cite{bruenger:06} given by
\bea
H_{KN}=J\sum_{\la ij\ra}\btau_i\cdot\btau_j + J_K\sum_i\btau_i\cdot\bS_i
\label{eq:hkn}
\eea
Here the kinetic hopping term in Eq.~(\ref{eq:KLU}) is replaced by an effective inter-site spin exchange $J=\frac{4t^2}{U}$ in the strong correlation limit of conduction electrons. The above  Hamiltonian is of the generalized 'Kondo-necklace' type (in 2D) originally studied for 1D chains with only the xy inter-site terms included~\cite{doniach:77}. It has, however, later been extended to higher dimension and including all components in the intra- and inter- site exchange  with possible uniaxial anisotropies for both~\cite{zhang:00,langari:06,thalmeier:07}. The model may also be viewed as an asymmetric bilayer Heisenberg model~\cite{kotov:98} with $\btau_i$ and $\bS_i$ spins residing on different layers and only the former coupled by inter-site exchange $J$. These are generic models to describe quantum phase transitions between a singlet ('Kondo') phase favored by the second term and an antiferromagnetically ordered phase favored by the first term. The transition occurs when the control parameter $J/J_K$  is larger than a the value  $(J/J_K)_c$ defining the quantum critical point (QCP). In 2D we have in the present isotropic model $(J/J_K)_c = 0.88$~\cite{langari:06,thalmeier:07}. Such transitions are frequently found in f-electron compounds where the control parameter may be varied by pressure or doping (i.e., chemical pressure). The above model allows to study the characteristic quantum critical behavior around the QCP disregarding the charge fluctuations. It has been investigated using numerical methods like Monte Carlo (MC) simulations~\cite{assaad:99,brenig:06}, exact diagonalization methods~\cite{zerec:06a} and dynamical mean field theory (DMFT) and also analytical methods like bond operator approach in mean field~\cite{zhang:00,langari:06,thalmeier:07} or hard-core boson treatment~\cite{rezania:08}. In this approach the Kondo  necklace Hamiltonian in Eq.~(\ref{eq:hkn}) is mapped to a  model of interacting singlet ($s$) and triplet ($t_\alpha, \alpha=x,y,z$) bosons by the bond-operator transformation~\cite{sachdev:90}. These bosons describe the singlet and triplet states $\ket{s}=s^{\dagger}\ket{0}$ and $\ket{t_\alpha}=t_{\alpha}^{\dagger}\ket{0}$ ($\alpha=x, y, z$) of the pair of spins ($\btau_i,\bS_i$) coupled by $J_K$ at every site i. In terms of bosonic operators the spins are given by
\bea
S_{i, \alpha}=\frac{1}{2}(s^{\dagger}_i t_{i, \alpha}+
t_{i, \alpha}^{\dagger} s_i -i \epsilon_{\alpha \beta \gamma} 
t_{i, \beta}^{\dagger}t_{i, \gamma}), \nonumber \\
\tau_{i, \alpha}=\frac{1}{2}(-s^{\dagger}_i t_{i, \alpha}-
t_{i, \alpha}^{\dagger} s_i -i \epsilon_{\alpha \beta \gamma} 
t_{i, \beta}^{\dagger}t_{i, \gamma}),
\label{eq:bondtrans}
\eea
where $\alpha, \beta, \gamma=x, y, z$
and $\epsilon_{\alpha\beta\gamma}$ is the totally antisymmetric tensor.
The conventional spin and boson commutation rules are fulfilled. The restriction
to physical states (only one boson per site) is expressed by the constraint
$s^{\dagger}_i s_i+\sum_{\alpha} t_{i, \alpha}^{\dagger} t_{i, \alpha}=1$ which may be implemented
either on the mean field level or by hard core boson technique. The former is chosen here.

\subsection{Ground state energy and triplon excitations}
\label{subsec:triplon}

In mean field approximation the bond operator transformation leads to
a bilinear bosonic Hamiltonian
\bea
H_{\text{KN}}^{\text{mf}}
&=&
\sum_{\bk,\alpha}(\frac{1}{2}J\bars^2\gamma_\bk + \frac{1}{4}+\mu)
t^\da_{\bk\alpha}t_{\bk\alpha}
\nonumber\\
&\phantom{=}&{}
+\sum_{\bk}\frac{1}{4}J\bars^2\gamma_\bk
(t^\da_{\bk\alpha}t^\da_{-\bk\alpha}+t_{\bk_\alpha}t_{-\bk\alpha})
\label{eq:hmf}
\eea
where $\mu$ is the chemical potential introduced to ensure the constraint and
$\bars=\la s\ra$ is the mean field singlet amplitude.  Furthermore we
define
$t_{\bk\alpha}=\frac{1}{\sqrt{N}}\sum_i\exp(i\bk\bR_n)t_{i\alpha}$ and
$\gamma_\bk=(\cos k_x +\cos k_y)$.  Here we restrict to the
nonmagnetic  phase where the triplet amplitude
$\bar{t}\equiv 0$.  In this case $\bars$ is always close to one.  This
Hamiltonian may be diagonalised by the Bogoliubov transformation
\bea
a_{\bk\alpha}&=&\cosh(\phi_ {\bk}) t_{\bk, \alpha} 
+\sinh(\phi_{\bk}) t_{-\bk\alpha}^{\dagger}, \nonumber \\
a_{-\bk\alpha}^{\dagger}&=&\sinh(\phi_{\bk})  t_{\bk, \alpha}
+\cosh(\phi_{\bk}) t_{-\bk, \alpha}^{\dagger},
\label{eq:bogol}
\eea
with 
\bea
\label{eq:functions}
\tanh(2\phi_\bk)&=&\frac{2f_\bk}{d_\bk}\no\\
f_\bk&=&\frac{1}{4}J\bars^2\gamma_{\bk}\\
d_\bk&=&\mu+\frac{1}{4}J_K+
\frac{1}{2}J\bar{s}^2\gamma_\bk\no
\eea
Diagonalization leads to the bosonic triplon Hamiltonian
\bea
H_{KN}^{mf}&=&E_0+\sum_\bk \sum_{\alpha=x,y,z} 
\omega_{\bk} a_{\bk, \alpha}^{\dagger}  a_{\bk, \alpha}\no\\
\frac{E_0}{N}&=&\Big(\mu (\bar{s}^2-1)-\frac{3}{4}J_K\bar{s}^2\Big)+
\frac{3}{2N}\sum_k 
\Big(\omega_{\bk}-d_{\bk}\Big)\no\\
\omega_{\bk}&=&\sqrt{d_\bk^2-4f_\bk^2}
\label{eq:htriplon}
\eea
where $\omega_{\bk}$ is the threefold degenerate ($\alpha =x,y,z$) triplon dispersion and $E_0(\mu,\bars)$ the ground state energy. 
Minimization of the latter leads to selfconsistency equations for the chemical potential $\mu$ and
singlet amplitude $\bars$ given by
\bea
\frac{3}{2N}\sum_{\bk}\frac{d_\bk}{\omega_\bk}
&=&\frac{5}{2}-\bar{s}^2\nonumber\\
\frac{3J}{2N}\sum_{\bk}\frac{d_\bk-2f_\bk}{\omega_{\bk}}
\gamma_\bk&=&\frac{3}{2}J_K-2\mu.
\label{eq:selfcons}
\eea
The smooth dependence of $\mu, \bars$ on the control parameter $J_K/J$ is shown in the inset of  Fig.~\ref{fig:Fig7}. It also 
extends continuously across the QCP into the antiferromagnetic region~\cite{thalmeier:07}.

The bond operator method also gives an explicit expression for the singlet-triplet gap which is defined as $\Delta_t=\omega_\bQ$ with 
$\bQ = (\pi,\pi)$~\cite{langari:06}. Using Eqs.~(\ref{eq:functions},\ref{eq:htriplon}) and setting $J=4t^2/U$ we obtain
\bea
\Delta_t= (\mu+\frac{1}{4}J_K)\bigl[1-2\frac{8t^2\bars^2}{U(\mu+\frac{1}{4})J_K}\bigr]^\frac{1}{2}
\label{eq:spingap}
\eea
This expression for the gap may be compared to the Kondo temperature scale $T^*$ from the susceptibility maximum obtained in FTLM finite cluster calculations.

%
\begin{figure}[Ht]
\centering
\includegraphics[width=75mm]{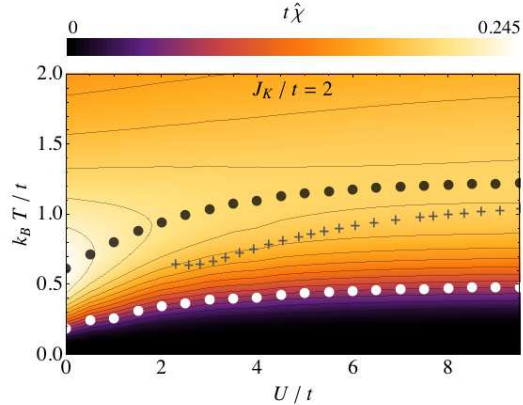}
\caption{Susceptibility in $(T,U)$ plane for fixed $J_K$. The Kondo temperature scale (black dots) increases first linearly 
with U and then reaches a plateau qualitatively similar to the specific heat result in Fig.~\ref{fig:Fig4}. For comparison we also show the temperature of the $\chi^{(3)}(T)$ maximum (white dots) and the triplon gap $\Delta_t$ (Eq.~(\ref{eq:spingap})) of the large-U Kondo necklace model (black crosses). }
\label{fig:Fig6}
\end{figure}
%

\subsection{The paramagnetic effective local moment}
\label{subsec:moment}

In the numerical calculation the central quantity in the correlated Kondo model is the local paramagnetic moment  \mul~   which is a direct measure of the singlet formation as function of $J_K/t$ and $U/t$. This also extends to the large U limit where the local moment
may be calculated analytically within bond operator approach. To gain a better understanding of the numerical results a comparison with the analytical method for large U is helpful. In spirit this is similar to the comparison of ED results for the $J_1-J_2$ Heisenberg model with analytical spin wave calculations~\cite{schmidt:11}. The elementary excitations here are, however, gapped triplon modes in the paramagnetic ('Kondo singlet ') regime rather than spin waves in the antiferromagnetic broken symmetry state.

First we transform the expression for \mul~ given in Eq.~(\ref{eq:muloc}) to bond operator basis. Using the defining relations in Eq.~(\ref{eq:bondtrans}) with $\theta_\alpha=\frac{1}{2}(\bt^\dagger\times\bt)_\alpha$ we derive the operator identity
%
%
%
\bea
\mu_{loc}^2=4\sum_\alpha\theta_\alpha^\dagger\theta_\alpha; \;\;
\la\mu_{loc}^2\ra=12\la\theta_x^\dagger\theta_x\ra
\label{eq:multheta}
\eea
where we used the spin space isotropy or degeneracy of triplon modes for the expectation value. Using the explicit expressions for $\theta_\alpha$, defining  $u_\bk=\cosh(\phi_ {\bk})$, $v_\bk=\sinh(\phi_ {\bk})$ and performing a Fourier transformation we obtain a concise form of the moment:
\bea
\frac{1}{6}\la\mu^2_{loc}\ra=\bigl(\frac{1}{N}\sum_\bk v_\bk^2\bigr)\bigl(\frac{1}{N}\sum_\bk u_\bk^2\bigr)
-\bigl(\frac{1}{N}\sum_\bk u_\bk v_\bk\bigr)^2
\nonumber\\
\label{eq:muluv}
\eea
The positiveness of the moment squared is ensured by Schwartz' inequality. Using Eq.~(\ref{eq:functions}) we can write
\bea
u_\bk^2&=&\frac{1}{2}\bigl(\frac{d_\bk}{\omega_\bk}+1\bigr);\;\;
v_\bk^2=\frac{1}{2}\bigl(\frac{d_\bk}{\omega_\bk}-1\bigr)\no\\
u_\bk v_\bk&=&\frac{f_\bk}{\omega_\bk};\;\;
u_\bk^2+v_\bk^2=\frac{d_\bk}{\omega_\bk}
\label{eq:uv}
\eea
Inserting these explicit expressions into Eq.~(\ref{eq:muluv}) we obtain the final result for the paramagnetic moment as
\bea
\la\mu^2_{loc}\ra=\frac{3}{2}\Bigl[ 
\frac{1}{N}\sum_\bk\bigl(\frac{d_\bk+2f_\bk}{d_\bk-2f_\bk}\bigr)^\frac{1}{2}
\frac{1}{N}\sum_\bk\bigl(\frac{d_\bk-2f_\bk}{d_\bk+2f_\bk}\bigr)^\frac{1}{2}
-1 \Bigr]
\nonumber\\
\label{eq:mulocbond}
\eea
where $d_\bk,f_\bk$ are defined in Eq.~(\ref{eq:functions}). From this closed expression the moment may be 
obtained by the 2D momentum integration.

It is worthwhile to consider the expression of $\la\bmu^2_{loc}\ra$ in the strong Kondo coupling limit $J/J_K\ll 1$ or $(4t^2/UJ_K\ll 1)$. 
We define $\epsilon_\bk =\frac{2f_\bk}{d_\bk}$ and use $\omega_\bk =d_\bk\sqrt{1-\epsilon_\bk^2}$. 
Expanding Eq.~(\ref{eq:muloc}) in terms of $\epsilon_\bk$ we get
\bea
\la\mu^2_{loc}\ra=\frac{3}{2}\Bigl[\frac{1}{N}\sum_\bk\epsilon_\bk^2 
-(\frac{1}{N}\sum_\bk\epsilon_\bk)^2\Bigr]
\eea
Which may be further evaluated to give
\bea
\la\mu^2_{loc}\ra=\frac{3}{8}\Bigl(\frac{\bars^2J}{\mu+\frac{1}{4}J_K}\Bigr)^2=
\frac{3}{8}\Bigl(\frac{4t^2\bars^2}{U(\mu+\frac{1}{4}J_K)}\Bigr)^2
\eea
Where in the last expression we replaced $J=\frac{4t^2}{U}$. This equation demonstrates that
the local moment is decreasing with increasing $J_K$ (note that $\mu$ and $\bars$ also depend on $J_K/J$ ) and 
with increasing correlation U. This is what the full numerical calculation of Eq.~(\ref{eq:mulocbond})  discussed below indeed confirms.

\subsection{Spin correlations and high temperature susceptibility}
\label{subsec:correlations}

To obtain a more detailed insight into the Kondo singlet formation and the influence of Coulomb correlations on it we have 
previously also calculated the evolution of on-site and next-neighbor spin correlations in the ED approach. It is desirable
to calculate them with bond operator approximation in large U limit for comparison with numerical results.
First we consider the on-site spin correlation function $S_{KS}(i)=\langle\btau_i\cdot\bS_i\rangle$ between localized and itinerant spins
and also the complementary partial local moments $\langle\btau_i^2\rangle$ and $\langle\bS_i^2\rangle$. 
Transforming to bond operator representation and using the isotropy we obtain
\bea
S_{KS}(i)&=&\frac{1}{4}\bigl[\la\mu^2_{loc}\ra-3\bars^2\la(t_{ix}+t_{ix}^\dagger)^2\ra\bigr]\\
\langle\btau_i^2\rangle = \langle\bS_i^2\rangle&=&
\frac{1}{4}\bigl[\la\mu^2_{loc}\ra+3\bars^2\la(t_{ix}+t_{ix}^\dagger)^2\ra\bigr]\no
\label{eq:sks1}
\eea
The equality $\langle\btau_i^2\rangle = \langle\bS_i^2\rangle$ is only valid in the localized  limit but does not hold in the original KLU model defined in Eq.~(\ref{eq:KLU}) for small U.
The site index i is suppressed i.f. because these local quantities are uniform. Expressing the triplet operators $t_x, t_x^\dagger$
 in terms of triplon eigenmodes by using Eq.~(\ref{eq:bogol}) this leads to 
 \bea
 \la(t_x+t_x^\dagger)^2\ra=\frac{1}{N}\sum_\bk(u_\bk+v_\bk)^2
 \eea
from Eq.~(\ref{eq:uv}) we finally obtain
\bea
S_{KS}&=&\frac{1}{4}\bigl[\la\mu^2_{loc}\ra-3\bars^2
\frac{1}{N}\sum_\bk\Bigl(\frac{d_\bk+2f_\bk}{d_\bk-2f_\bk}\Bigr)^\frac{1}{2}\bigr]\no\\
\langle\btau_i^2\rangle = \langle\bS_i^2\rangle&=&\frac{1}{4}\bigl[\la\mu^2_{loc}\ra+3\bars^2
\frac{1}{N}\sum_\bk\Bigl(\frac{d_\bk+2f_\bk}{d_\bk-2f_\bk}\Bigr)^\frac{1}{2}\bigr]
\label{eq:sks2}
\eea
If we take the sum of these expression according to Eq.~(\ref{eq:muloc}) one obtains again the total local moment.

The total moment in the large-$U$ (localized spin) limit calculated in
Sec.~\ref{subsec:moment} is a zero temperature quantity determined by
quantum fluctuations in the ground state.  On the other hand the
effective moment of a localized spin system is obtained from the high
temperature behavior of the susceptibility.  The latter may be calculated 
from a high temperature expansion.
In this section we want to investigate how these effective moments are
related and how well the high temperature expansion agrees with the
FTLM results in the large-$U$ limit.
%
\begin{figure}
\centering
\includegraphics[width=75mm]{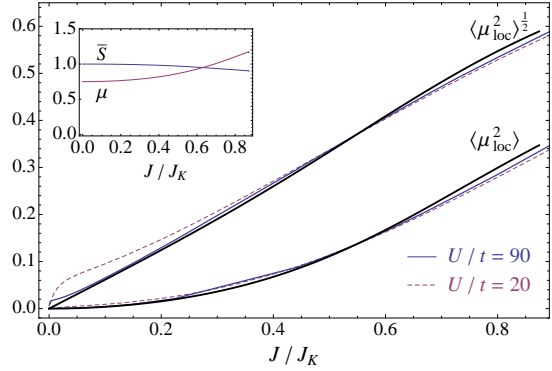}
\caption{Paramagnetic local moment and its square  in the large U-limit as function of $J/J_K$ where $J=4t^2/U$.
Thick line gives the result of bond operator theory. Thin and broken line are ED results for the eight site cluster in the large U limit. The good  agreement suggest that finite size effects for the local moments are small. The inset shows the dependence of singlet amplitude
$\bar{s}$ and chemical potential $\mu$ in mean field bond operator theory (Sec.~\ref{sec:bondop}). }
\label{fig:Fig7}
\end{figure}
%
First we give the high temperature expansion for the total local
moment at finite temperature $T$:
\begin{equation}
    \la\mu_{\text{loc}}^{2}\ra(T)
     =
    \left\langle\bS_{i}^{2}\right\rangle_T+
    \left\langle\btau_{i}^{2}\right\rangle_T+
    2\left\langle\bS_{i}\cdot\btau_{i}\right\rangle_T
    \label{eqn:mulocT}
\end{equation}
where
\[
\left\langle A\right\rangle_T
=
\frac{1}{\cal Z}
\mathop{\rm tr}\left(A{\rm e}^{-\beta H_{\text{KN}}}\right),
\quad
{\cal Z}=\mathop{\rm tr}\left({\rm e}^{-\beta H_{\text{KN}}}\right)
\]
is the thermal expectation value of $A$ formed with the Kondo necklace
Hamiltonian.  Expanding the statistical operator  for large $k_BT$ we obtain
%
\begin{eqnarray}
    \la\mu_{\text{loc}}^{2}\ra(T)
    &=&
    2S(S+1)
    \left[
    1-\frac{S(S+1)}{3}\frac{J_{\text K}}{k_{\text B}T}
    +{\cal O}\left(\frac{J_{\text K}}{k_{\text B}T}\right)^{2}
    \right]
    \nonumber
    \\
    &\equiv&
    \hat{\mu}_{\text{loc}}^{2}(T)
    +{\cal O}\left(\frac{J_{\text K}}{k_{\text B}T}\right)^{2}.
    \label{eq:muloccurie}
\end{eqnarray}
The prefactor is the paramagnetic moment of uncorrelated spins, where
the factor $2$ is due to the presence of two spins $\bm\tau$ and $\bm 
S$ per site with $\tau=S=1/2$.  The parenthesis gives the first order
high temperature correction of the local moment which depends only on
the local Kondo exchange $J_{\text K}$.  Likewise we can calculate the high
temperature uniform susceptibility, which was defined previously as
\begin{eqnarray}
    \hch(T)&=&\frac{1}{N}
    \frac{1}{k_{\text 
    B}T}\left\langle\left(S_{z}^{\text{tot}}\right)^{2}\right\rangle_T
    \label{eq:chi2}
 \end{eqnarray}
where the same g-factor has been assumed for both $\btau$- and
$\bS$-spins.  Evaluating the thermal average in high temperature
approximation as before and using $\tau=S$ one obtains for the high temperature susceptibility
%
%
%
\begin{eqnarray}
   \label{eq:suscw}    
    \hch(T)
    &=&
    \frac{1}{3}
    \frac{1}{k_{\text B}T}\times
    \\
   && 2S(S+1)
    \left[
    1-\frac{S(S+1)}{3}\frac{J_{\text K}}{k_{\text B}T}
    -\frac{S(S+1)}{3}\frac{zJ}{k_{\text B}T}
    \right]
    \nonumber\\
    &\approx&
    \frac{1}{3}
    \frac{1}{k_{\text B}T}
    \times\la\mu_{\text{loc}}^{2}\ra(T)
    \left[
    1-\frac{S(S+1)}{3}\frac{zJ}{k_{\text B}T}
    \right].
    \nonumber   
\end{eqnarray}  
Formally, we can regard the last expression as the first two terms of 
a high-temperature expansion of a Curie-Weiss type susceptibility
\begin{equation}
    \hch(T)=\frac{\hch_{0}(T)}{1+\Theta/T}
\end{equation}
with an effective noninteracting Curie susceptibility $\hch_{0}(T)$ and a
Weiss temperature $\Theta$ given, respectively by
\begin{eqnarray}
    \hch_0(T)
    &=&
    \frac{1}{3}\frac{1}{k_{\text B}T}
    \times
    \hat\mu_{\text{loc}}^{2}(T),
    \nonumber\\
    k_{\text B}\Theta
    &=&
    \frac{S(S+1)}{3}zJ 
    \label{eq:cwtemp}
\end{eqnarray}
i.e.  $ k_{\text B}\Theta = J$ for  $S=\frac{1}{2}, z=4$. 
We write the high-temperature behavior of the susceptibility $\hch(T)$
in this suggestive form to stress that the temperature-dependent
effective moment screening is due to the Kondo coupling $J_{\text K}$,
and the Weiss temperature $\Theta$ due to the effective magnetic
intersite exchange $J$.  One may conjecture that this expression may be
valid beyond its formal expansion regime.  This may be checked by
comparing Eq.~(\ref{eq:suscw}) with the unbiased FTLM results for
$\hch(T)$ and $\hat{\mu}_{\text{loc}}^{2}(T)$.
%
\begin{figure}
\centering
\includegraphics[width=75mm]{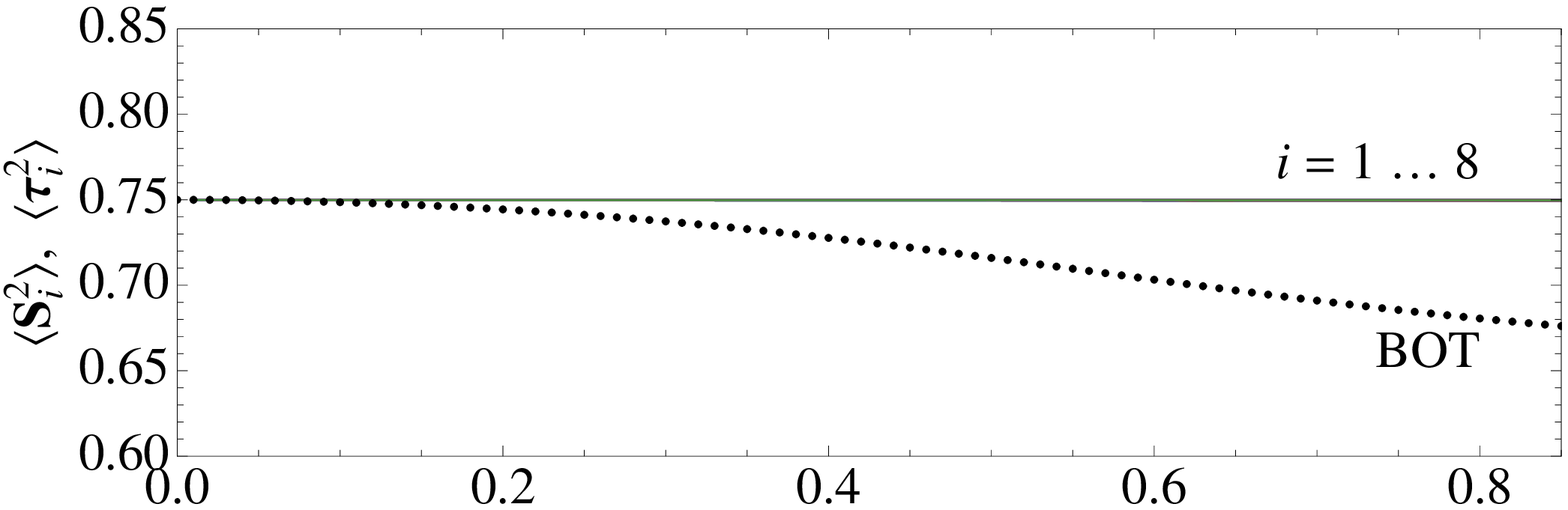}
\includegraphics[width=75mm]{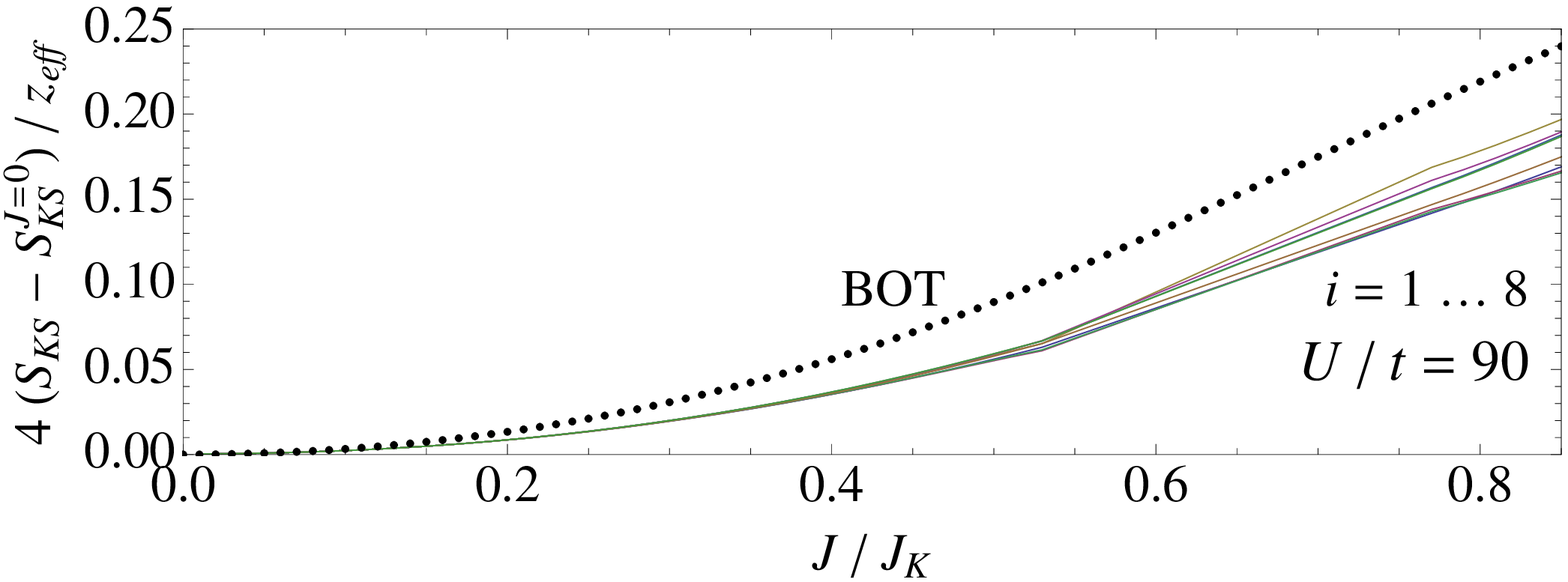}
\caption{Comparison of partial moments (top) and on-site AF correlations  $S_{KS}(i)=\langle\btau_i\cdot\bS_i\rangle$ (bottom)
from ED (full lines) and bond operator theory (BOT) (dotted line). The results for $S_{KS}$ have been scaled by the effective coordination number of a given cluster site ($z_{eff} =1..4$; for BOT $z_{eff}=4$). The deviations in both quantities have different
signs and almost cancel in the total local moment \mul (Fig.~\ref{fig:Fig7}).}
\label{fig:Fig8}
\end{figure}
%

\section{Comparison with numerical results}
\label{sec:comparison}

It is an interesting question whether the previous analytical mean field results for the large U limit can be qualitatively 
compared to the numerical ED results for finite tiles in Sec.~\ref{sec:discussion}. The most obvious quantity to
 check this is $\la\mu_{\text{loc}}^{2}\ra$.
The comparison is shown in Fig.~\ref{fig:Fig7}. The inset of this figure gives the dependence of $\mu$ and $\bars$ on $J/J_K$ as obtained from the self-consistent solution of Eq.~(\ref{eq:selfcons}) almost  up to the quantum critical  value $(J/J_K)_c=0.88$ where the singlet-triplet spin gap vanishes and AF order would set in. We stay in the paramagnetic parameter range throughout this work. The AF region of the 2D Kondo necklace region has been explored in Refs.~\cite{langari:06, thalmeier:07}. 

The main Fig.~\ref{fig:Fig7} shows the dependence of the local moment $\la\mu_{loc}^2\ra^\frac{1}{2}$ and its square on $J/J_K$ from bond-operator mean field theory (full black line) and from numerical results for the eight site cluster model in the large $U$ limit with $J=4t^2/U$. Here large $U$ means large as compared to the tight binding bandwidth $W=2zt =8t$. The numerical results for various $U/t$ in this limit are given as dashed lines. 
The agreement of analytical and numerical results is surprisingly good. This proves two points: Firstly the mean-field bond operator technique gives reliable on-site properties in the ground state. This is in agreement with the observation~\cite{thalmeier:07} that values for $(J/J_K)_c$ that determine the quantum critical point between Kondo singlet and AF phase are well reproduced by that theory. Secondly the agreement points to the fact that the numerical finite size effects on a local quantity like $\la\mu_{\text{loc}}^{2}\ra$ are apparently quite moderate. Noticeable deviations in the numerical and analytical results appear at larger $J/J_K$ when the quantum critical point to AF order is approached or when U/t becomes too small.  

%
\begin{figure}
\centering
\includegraphics[width=75mm]{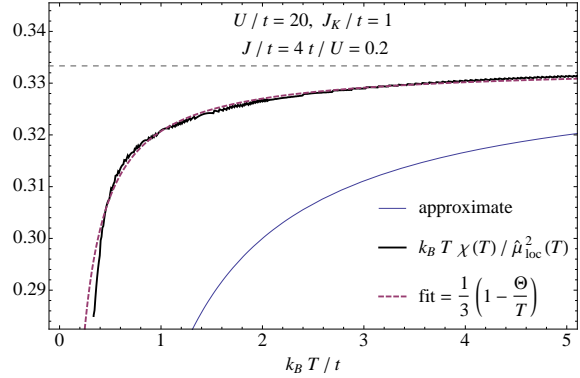}
\caption{Susceptibility and effective moment from FTLM (upper full line). Dashed line is obtained from high temperature  expansion formula using $k_B\Theta = 0.0379t$ as a fit parameter. It is considerably smaller than in Eq.~(\ref{eq:suscw}) due to reduced coordination number in finite clusters. Lower full lines from high temperature expansion  with $\Theta$ given by Eq.(\ref{eq:cwtemp}) using the approximate expression for 
$\chi(T)$ in Eq.~(\ref{eq:suscw}).}
\label{fig:Fig9}
\end{figure}
%
The comparison may also be made for the on-site spin correlation functions $S_{KS}(i)$ and the partial moments $\langle\btau^2\rangle$ and $\langle\bS^2\rangle$ which are equal in the large U limit. The latter is shown in Fig.~\ref{fig:Fig8}a with numerical results from ED giving the proper local moment value $3/4$ and the results from bond operator theory lying below. Their difference increases with increasing $J$, i.e. decreasing $U$. On the other hand the on-site AF Kondo correlation presented in  Fig.~\ref{fig:Fig8}b shows the analytical result lying above the numerical ED value by a similar amount. The latter depends on the effective coordination number of the considered site in the finite cluster. The total moment \mul~ is the sum of these individual contributions and because of the opposite sign of the differences the add up to zero approximately. This explains why numerical ED and analytical bond operator results for  \mul~ in Fig.~\ref{fig:Fig7} agree so well despite the small cluster size. Including the previous results~\cite{langari:06,thalmeier:07} on quantum critical properties of the Kondo necklace model we can  conclude that the energetics and local correlations are well described by the mean field bond operator method. However, it should be noted that the description of  inter-site correlations and their U dependence is well beyond this approximation.

Finally we come to the high temperature expansion for local moment and susceptibility in the large U limit and its comparison to the FTLM results. According to Eq.~(\ref{eq:suscw}) the ratio $T\chi(T)/\hat{\mu}^2_{\text{loc}}(T)$ should be proportional to the Curie Weiss factor in this equation. This comparison is shown in Fig.~\ref{fig:Fig9}. It is seen that the result of the high temperature expansion lies considerably below the curve obtained from FTLM. Part of this discrepancy may be due to the fact that in FTLM the effective (average) coordination number in the eight site tile  is smaller than $z=4$. Therefore, in order to compare with high temperature expansion we may consider an effective Curie Weiss temperature $\Theta_{\text{eff}}$ as fit parameter in Eq.~(\ref{eq:suscw}). Then the temperature dependence of $\chi(T)$ from FTLM is well reproduced by the form of the high temperature expansion results in Fig.~\ref{fig:Fig9}.

\section{Summary and conclusion}
\label{sec:summary}

In this work we investigated the local moment screening and spin correlations and in particular thermodynamic properties of the correlated Kondo lattice or Kondo-Hubbard model. We have focused primarily on the systematic dependence of local moment, spin correlations and susceptibility on the control parameters $J_K $ and $U$ and secondly on the U-dependence of the characteristic temperature scale $T^*$. We used  unbiased  and exact  numerical techniques like ED and FTLM for small tiles for the whole range of $U/t$ and $J_K/t$ as well as approximate analytical bond-operator method for the large U-limit of the extended Kondo necklace model which contains only localized spins. In this limit both methods give excellent agreement on the dependence of the Kondo-screened total paramagnetic local moment (Fig.~\ref{fig:Fig7}) on $U$ or $J=4t^2/U$. Our ED and FTLM investigation may also be of particular relevance for finite nano- clusters of Kondo  atoms adsorbed on surfaces.

A first central result obtained  for smaller U and $J_K$ is a clear non-monotonic behaviour of \mul~ on correlation strength is observed in the ED results. This non-monotonic U-dependence is also observed in the on-site and inter-site spin correlations where the latter are of mixed RKKY and superexchange character. It is the result of competing effects of conduction electron localization by U and local singlet formation due to $J_K$.

The second central conclusion concerns the dependence of the Kondo temperature scale $T^*$ on correlation strength U for which controversial results have been reported previously. Our FTLM susceptibility and the analytical results presented in Fig.~\ref{fig:Fig6} show that it increases monotonically with U and reaches a plateau in the large U limit. In this limit the Kondo scale corresponds to the spin-gap for triplon excitations at the AF wave vector $\bQ=(\pi,\pi)$.

Furthermore a high temperature expansion for the susceptibility of the Kondo necklace model leads to an effective Curie-Weiss type expression where the local susceptibility is modified only by the Kondo term and the Curie-Weiss temperature is only due to the superexchange term. The resulting  temperature dependence is similar to that of FTLM results in the large U limit and if the effective Curie-Weiss temperature is used as a fit parameter a quantitative agreement  over large temperature region is obtained.

\clearpage
\section*{References}

\bibliography{corrkon}

\providecommand{\newblock}{}
\begin{thebibliography}{10}
\expandafter\ifx\csname url\endcsname\relax
  \def\url#1{{\tt #1}}\fi
\expandafter\ifx\csname urlprefix\endcsname\relax\def\urlprefix{URL }\fi
\providecommand{\eprint}[2][]{\url{#2}}

\bibitem{hewson:93}
Hewson A 1993 {\em The Kondo problem to heavy fermions\/} (Cambridge University
  Press)

\bibitem{newns:87}
Newns D~M and Read N 1987 {\em Adv. Phys.\/} {\bf 36} 799

\bibitem{bickers:87}
Bickers N~E 1987 {\em Rev. Mod. Phys.\/} {\bf 59} 845

\bibitem{zwicknagl:92}
Zwicknagl G 1992 {\em Adv. Phys.\/} {\bf 41} 203

\bibitem{khaliullin:95}
Khaliullin G and Fulde P 1995 {\em Phys. Rev. B\/} {\bf 52} 9514

\bibitem{neef:03}
Neef M, Tornow S, Zevin V and Zwicknagl G 2003 {\em Phys. Rev. B\/} {\bf 68}
  035114

\bibitem{schork:94}
Schork T and Fulde P 1994 {\em Phys. Rev. B\/} {\bf 50} 1345

\bibitem{li:95}
Li Y~M 1995 {\em Phys. Rev. B\/} {\bf 52} R6979

\bibitem{takayama:98}
Takayama R and Sakai O 1998 {\em J. Phys. Soc. Jpn.\/} {\bf 67} 1844

\bibitem{itai:96}
Itai K and Fazekas P 1996 {\em Phys. Rev. B\/} {\bf 54} R752

\bibitem{shibata:96}
Shibata N, Nishino T, Ueda K and Ishii C 1996 {\em Phys. Rev. B\/} {\bf 53}
  R8828

\bibitem{feldbacher:02}
Feldbacher M, Jurecka C, Assaad F~F and Brenig W 2002 {\em Phys. Rev. B\/} {\bf
  66} 045103

\bibitem{jurecka:01}
Jurecka C and Brenig W 2001 {\em Phys. Rev. B\/} {\bf 64} 092406

\bibitem{zerec:06a}
Zerec I, Schmidt B and Thalmeier P 2006 {\em Physica B\/} {\bf 378-380} 702

\bibitem{zerec:06b}
Zerec I, Schmidt B and Thalmeier P 2006 {\em Phys. Rev. B\/} {\bf 73} 245108

\bibitem{igarashi:95}
Igarashi J, Tonegawa T, Kaburagi M and Fulde P 1995 {\em Phys. Rev. B\/} {\bf
  51} 5814

\bibitem{zhang:00}
Zhang G~M, Gu Q and Yu L 2000 {\em Phys. Rev. B\/} {\bf 62} 69

\bibitem{langari:06}
Langari A and Thalmeier P 2006 {\em Phys. Rev. B\/} {\bf 74} 024431

\bibitem{thalmeier:07}
Thalmeier P and Langari A 2007 {\em Phys. Rev. B\/} {\bf 75} 174426

\bibitem{neel:08}
N\'eel N, Kr\"oger J, Berndt R, Wehling T~O, Lichtenstein A~I and Katsnelson
  M~I 2008 {\em Phys. Rev. Lett.\/} {\bf 101} 266803

\bibitem{neel:11}
N\'eel N, Berndt R, Kr\"oger J, Wehling T~O, Lichtenstein A~I and Katsnelson
  M~I 2011 Two-site kondo effect in atomic chains {arXiv}:1105.3301

\bibitem{schmidt:11}
Schmidt B, Siahatgar M and Thalmeier P 2011 {\em Phys. Rev. B\/} {\bf 83}
  075123

\bibitem{tsunetsugu:92}
Tsunetsugu H, Hatsugai Y, Ueda K and Sigrist M 1992 {\em Phys. Rev. B\/} {\bf
  46} 3175

\bibitem{jaklic:00}
Jaklic J and Prelovsek P 2000 {\em Adv. Phys.\/} {\bf 49} 1--92

\bibitem{affleck:09}
Affleck I 2009 The kondo screening cloud: what it is and how to observe it
  {arXiv}:0911.2209

\bibitem{schmidt:05}
Schmidt B and Thalmeier P 2005 {\em Physica B\/} {\bf 359-361} 1387--1390

\bibitem{ramirez:92}
Ramirez A~P, Coleman P, Chandra P, Br\"uck E, Menovsky A~A, Fisk Z and Bucher E
  1992 {\em Phys. Rev. Lett.\/} {\bf 68} 2680

\bibitem{bruenger:06}
Br\"unger C and Assaad F~F 2006 {\em Phys. Rev. B\/} {\bf 74} 205107

\bibitem{doniach:77}
Doniach S 1977 {\em Physica B+C\/} {\bf 91} 231 -- 234

\bibitem{kotov:98}
Kotov V~N, Sushkov O, Weihong Z and Oitmaa J 1998 {\em Phys. Rev. Lett.\/} {\bf
  80} 5790

\bibitem{assaad:99}
Assaad F~F 1999 {\em Phys. Rev. Lett.\/} {\bf 83} 796

\bibitem{brenig:06}
Brenig W 2006 {\em Phys. Rev. B\/} {\bf 73} 104450

\bibitem{rezania:08}
Rezania H, Langari A and Thalmeier P 2008 {\em Phys. Rev. B\/} {\bf 77} 094438

\bibitem{sachdev:90}
Sachdev S and Bhatt R~N 1990 {\em Phys. Rev. B\/} {\bf 41} 9323

\end{thebibliography}

\end{document}